\documentclass[11pt,a4paper]{article}
\pdfoutput=1

\usepackage{bm}
\usepackage{amsmath,amssymb}
\usepackage{cite}

\usepackage{graphicx}
\usepackage{color}
\usepackage{upgreek}

\usepackage{hyperref} 
\hypersetup{colorlinks=true, citecolor=blue, filecolor=black, linkcolor=blue, urlcolor=blue, pdfpagemode=UseNone}

\numberwithin{figure}{section}
\numberwithin{equation}{section}

\newcommand{\be}{\begin{equation}}
\newcommand{\ee}{\end{equation}}
\newcommand{\bea}{\begin{eqnarray}}
\newcommand{\eea}{\end{eqnarray}}
\def\beal#1\eeal{\begin{align}#1\end{align}}   
\def\besp#1\eesp{\begin{multline}#1\end{multline}} 
\newcommand{\nn}{\nonumber}

\newcommand{\TRM}[1]{#1}
\newcommand{\new}[1]{#1}

\newcommand\ie{\textit{i.e.}\ }
\newcommand\eg{\textit{e.g.}\ }
\newcommand\cf{\textit{cf.}\ }

\newcommand{\etc}{{\it etc.}\ }
\newcommand{\viz}{{\it viz.}\ }
\newcommand{\half}{\tfrac{1}{2}}

\newcommand{\DeWitt}{DeWitt:1992cy}

\newcommand{\str}{\text{str}}
\newcommand{\sdet}{\text{sdet}}
\newcommand{\cu}[1]{\!#1\!}
\def\one{\hbox{1\kern-.8mm l}}

\begin{document}

\begin{titlepage}

\begin{center}
{\huge \bf Parisi-Sourlas supergravity} 


\end{center}
\vskip1cm


\begin{center}
{\bf Matthew Kellett and Tim R. Morris}
\end{center}

\begin{center}
{\it STAG Research Centre \& Department of Physics and Astronomy,\\  University of Southampton,
Highfield, Southampton, SO17 1BJ, U.K.}\\
\vspace*{0.3cm}
{\tt  M.P.Kellett@soton.ac.uk, T.R.Morris@soton.ac.uk}
\end{center}

\abstract{A manifestly diffeomorphism invariant exact renormalization group requires extra diffeomorphism invariant ultraviolet regularisation at some effective cutoff scale $\Lambda$. This motivates construction of a `Parisi-Sourlas' supergravity, in analogy with the gauge theory case,  where the superpartner fields have the wrong spin-statistics such that they can become Pauli-Villars regulator fields after spontaneous symmetry breaking. We show that in contrast to gauge theory, the free theory around flat space is already non-trivial and in a sense already displays some spontaneous symmetry breaking. We show that the fluctuating fields form multiplets whose mass matrices imply that the fields propagate into each other not only with the expected $1/p^2$ but also through propagators with improved ultraviolet properties, namely $1/p^4$ and $1/p^6$, despite the fact that the action contains a maximum of two space-time derivatives.}


\end{titlepage}

\tableofcontents


\section{Introduction and Motivation}
\label{sec:intro}

The renormalization group (RG) structure of quantum gravity is surely of importance,  see \eg \cite{Stelle:1976gc,Adler:1982ri,Weinberg:1980,Donoghue:1994dn,Reuter:1996,Ambjorn:2012jv,Morris:2018mhd,Morris:2018axr,Kellett:2018loq,first,secondconf,second,Loll:2019rdj,Bonanno:2020bil}, and central to this is the r\^ole of diffeomorphism invariance. In ref. \cite{Morris:2016nda} the first step was taken in combining these two properties transparently, by developing a manifestly diffeomorphism invariant Wilsonian exact RG\footnote{The name ``exact RG'' was introduced  in both refs. \cite{Wilson:1973,Wegner:1972ih}, for their continuum versions of the Wilsonian RG.} for gravity. Such a framework should allow both conceptual and computational advances. On the one hand it would allow computations to be done whilst keeping exact diffeomorphism invariance at every stage, \ie without gauge fixing, and on the other hand, these computations should be possible without first choosing the space-time manifold, in particular without introducing a separate background metric dependence. Indeed in ref. \cite{Morris:2016nda} such a framework was developed at the classical level where these properties were shown to hold. 

However in order to compute quantum corrections, extra ultraviolet regularisation has to be incorporated into the exact RG \cite{Morris:2016nda} so that the integration is properly cut off in some diffeomorphism invariant way at the effective cutoff scale $\Lambda$. 

In developments over a period of years this problem was solved for gauge field theory  \cite{Morris:1995he,Morris:1998kz,Arnone:2005vd,Morris:2006in,Rosten:2008zp,Morris:1999px,Morris:2000fs,Morris:2000jj,Arnone:2000bv, Arnone:2000qd,Arnone:2001iy,Arnone:2002yh,Arnone:2002qi,Arnone:2002fa,Arnone:2003pa,Arnone:2002cs,Arnone:2005fb,Arnone:2002fb,Gatti:2002kc,Morris:2005tv,Rosten:2004aw,Rosten:2005qs,Rosten:2005ka,Rosten:2005ep,Rosten:2006tk,Rosten:2006qx,Rosten:2006pd}, where it was proved to work to all orders in perturbation theory. (For a short summary see ref. \cite{Morris:2016nda}, and for reviews and further advances see refs. \cite{Arnone:2006ie,Rosten:2010vm,Rosten:2011ty,Falls:2017nnu}.\footnote{In particular in ref. \cite{Falls:2017nnu} the construction was generalised to curved backgrounds and used to compute the gauge field conformal anomaly, without gauge fixing.}) 
In gauge theory, this extra regularisation is provided by generalising the gauge group from $SU(N)$ to $SU(N|N)$ and then spontaneously breaking the fermionic gauge fields at the effective cutoff scale $\Lambda$. The resulting massive fields behave as gauge invariant Pauli-Villars fields with masses set by $\Lambda$ and interactions that are naturally incorporated into the flow equation, in such a way that they continue to regulate for all scales $\Lambda$  \cite{Morris:1998kz,Morris:1999px,Morris:2000fs,Morris:2000jj}. The reason these provide the needed extra regularisation can be understood as follows. The extra structure introduces as many wrong-statistics fermionic fields as there are bosonic degrees of freedom.\footnote{Actually for the counting to work exactly at finite $N$, it is first necessary to extend the group to $U(N|N)$ after which one sees that two vector bosons decouple \cite{Arnone:2001iy}.} 
For the gauge fields themselves, the original gauge field $A^1_\mu$ is joined by a copy gauge field $A^2_\mu$ (with wrong-sign kinetic term but which decouples in the continuum limit  in dimensions $D\le4$ \cite{Arnone:2001iy})
and a complex pair of fermionic gauge fields $B_\mu, \bar{B}_\mu$.
At high energies these degrees of freedom cancel each other, as happens with Parisi-Sourlas supersymmetry \cite{Parisi:1979ka}, at least sufficiently that\new{,} together with appropriately chosen covariant cutoff functions\new{,} the theory is then regularised to all orders in perturbation theory \cite{Arnone:2000bv, Arnone:2000qd,Arnone:2001iy}.

Given the developments just described it is natural to conjecture that the extra regularisation for gravity can be incorporated by introducing wrong-statistics fermionic components to the metric in a way that extends the diffeomorphism invariance along fermionic directions \cite{Morris:2016nda}.\footnote{Manifestly diffeomorphism invariant exact RGs are proposed  in \cite{Wetterich:2016ewc,Falls:2020tmj} that avoid introducing Pauli-Villars fields.} Working in Euclidean signature (so that the Wilsonian RG makes sense), we are therefore naturally led to consider extending the coordinates themselves to
\be 
\label{supercoords}
x^A = (x^\mu,\theta^a)\,,
\ee
such that the $D$-dimensional bosonic coordinates, $x^\mu$, are supplemented by $D$-dimensional real fermionic coordinates, $\theta^a$. Note that unlike for supergravity \cite{Freedman_1976,Gates:1983nr}, we want (the associated vector bundle to)  the Grassmann  $\theta^a$ to be vectorial in their own separate $D$-dimensional space rather than be spinorial under the (bosonic) Lorentz group.  This is so that superfields of the $\theta^a$ contain  wrong-statistics fermionic fields whose interactions mimic as closely as possible the bosonic fields,
thus implementing Parisi-Sourlas-type cancellations \cite{Parisi:1979ka} in a similar way to that just described for gauge theory.
Writing the invariant interval as 
\be 
\label{rough}
ds^2 \sim dx^A g_{AB}\, dx^B
\ee
(the precise definition will be given later),
we have introduced $D^2$ wrong-statistics fermionic degrees of freedom $g_{\mu a}= g_{a\mu}$, the right number to cancel the $D^2$ bosonic degrees freedom, namely the $D(D\cu+1)/2$ degrees of freedom in the original metric $g_{\mu\nu}$ and the $D(D\cu-1)/2$ bosonic degrees of freedom in the antisymmetric components $g_{ab}$ \cite{Morris:2016nda}.

We are thus led to consider a novel type of supergravity, which we might reasonably christen \emph{Parisi-Sourlas supergravity}. Fortunately, very general supermanifolds have been extensively developed in ref. \cite{DeWitt:1992cy}, and our construction will build on this. Obviously, for the construction to be successful, we need to verify that it does actually provide the desired cancellation of quantum corrections. The first step, which we take in this paper, is to understand more carefully the propagating degrees of freedom around a flat background supermetric. After suitable supercoordinate transformations, the latter must take the form:
\be
\label{flatg}
g_{AB} = \bar{\delta}_{AB}=\begin{pmatrix}
\delta_{\mu\nu} & 0\\
0 & \epsilon_{ab}
\end{pmatrix}\,,
\ee
where we write the flat metric in the fermionic directions as the constant antisymmetric $D\cu\times D$ matrix  $\epsilon_{ab}$. For the metric to be non-singular,  $\epsilon_{ab}$ must be  invertible. The dimension $D$ must therefore be even. By supercoordinate transformations we can (and will from now on) set
\be 
\label{deteps}
\det\epsilon = 1\,.
\ee

To understand what new degrees of freedom have been introduced (\ie over and above the graviton), we decompose the superfields into their component fields and analyse their transformation properties under linearised superdiffeomorphisms. After appropriate gauge fixing, we isolate the propagating degrees of freedom, and by diagonalising their kinetic terms, determine whether they have the right sign, or wrong sign (and thus are ghost-like).

We will need these extra propagating degrees of freedom to decouple at energies much lower than $\Lambda$. In gauge theory this is achieved by incorporating a $U(N|N)$ `Higgs' superfield,  which gains an expectation value of magnitude $\Lambda$, spontaneously breaking the fermionic directions and providing $B_\mu, \bar{B}_\mu$ with masses  \cite{Morris:2000fs,Morris:2000jj}. We therefore expect to have to introduce some analogous symmetry breaking, but we leave this step to a future paper.

In fact 
 we already find that around the background metric \eqref{flatg}, there is a sense in which some spontaneous symmetry breaking takes place. The kinetic terms are diagonalised only with the help of an arbitrary mass scale $M$. These fields are then seen to have mass terms proportional to $M$. However thanks to signs in the kinetic terms, the mass matrices do not behave in the normal way. Instead expansion in $M^2/p^2$ terminates after a few powers ($p$ being momentum). This behaviour is explained by working in the alternative basis in which $M$ does not appear. Here the fields are seen to propagate into each other via propagators with a fixed power $(1/p^2)^n$. Despite the fact that the action for the kinetic terms has a maximum of two space-time derivatives, the powers involve not only the expected $n\cu=1$, but also higher powers $n\cu=2,3$, these latter propagators thus having improved ultraviolet behaviour.
  
The structure of the paper is as follows. In sec. \ref{sec:review}, following closely ref. \cite{DeWitt:1992cy} we review the notation and key formulae we will need, in particular giving the precise definition of \eqref{rough}. In sec. \ref{sec:setup} we set up the action and expansion around the flat background \eqref{flatg}. In sec. \ref{sec:decomposition} we introduce the mass-scale $M$ and decompose superfields into components, with the help of the Hodge dual. In sec. \ref{sec:gf} we fix the gauge first by algebraic elimination, and then to a radiation gauge. Then finally in sec. \ref{sec:props} we are ready to analyse the propagating degrees of freedom and their properties. In sec. \ref{sec:discussion} we summarise and draw our conclusions.
 

\section{Supermanifolds: a review}
\label{sec:review}


We collect together here the basic material we will need to formulate Parisi-Sourlas supergravity.
We will mostly use notation, nomenclature and definitions from ref. \cite{DeWitt:1992cy}. 
For the moment we work in $D$ dimensions, although after we expand in component fields in sec. \ref{sec:decomposition}, it will be useful to specialise the physically interesting case of $D\cu=4$. Although ref. \cite{DeWitt:1992cy} allows for different numbers of fermionic and bosonic coordinates we want the same number for reasons already explained, thus we work on a superspace $\mathbb{R}_c^D\times \mathbb{R}_a^D$, and with indices $A,B,C,\dots$, such that $A=(\alpha,a)$ \etc  
The index $\alpha$ labels  c-type (commuting / bosonic) Euclidean coordinates, and $a$ labels a-type (anticommuting / fermionic) partners. 

A ``c-type" supervector \textbf{X}  has c-numbers in the first $D$ places and a-numbers in the last $D$ places (with respect to a standard basis). Similarly, an ``a-type" supervector has a-numbers in the first $D$ places and c-numbers in the last $D$ places.
In addition, we take indices to be on the left or right, as well as up or down \cite{DeWitt:1992cy}. This denotes slightly different transformation properties (see below). We use notations such as
\be
(-1)^A,\quad (-1)^{\textbf{X}}, \quad (-1)^{\textbf{X}A}
\ee
In this notation, $A$ is not meant to be read as an index (in the sense of the Einstein convention) but as a label which is 0 for $A=\alpha$ and 1 for $A=a$. For the supervector, we say $(-1)^{\textbf{X}}=1$ for $\textbf{X}$ c-type, and $(-1)^{\textbf{X}}=-1$ for $\textbf{X}$ a-type. In general, when an object or index appears in a power of $(-1)$, this is to be read as the value of its $\mathbb{Z}_2$ Grassmann grading (either 1 or 0).
If $\textbf{X}$ is c-type or a-type, it is said to be a ``pure" supervector. The above definitions only apply for pure supervectors, but for our purposes we can extend formulae linearly since all supervectors can be expressed (uniquely) as the sum of a c-type and an a-type supervector.
When dealing with contractions, we take the usual convention that we can only contract up indices with down indices, however also that the ``natural" contraction is between adjacent indices (with no object, index, supervector or otherwise between them) or an index-dependent sign will appear.


A supervector space is defined in the same way as a vector space, except that it is a space over $\mathbb{R}_c^D\times \mathbb{R}_a^D$ and with left/right multiplication in general being different maps. In general we use what DeWitt \cite{DeWitt:1992cy} calls a ``standard basis" $\{{}_A\textbf{e}\}$, which has the following behaviour under complex conjugation:
\be
\label{standard}
{}_A\textbf{e}^* = (-1)^A{}_A\textbf{e}
\ee
This means that a ``real" supervector $\textbf{X}=X^A{}_A\textbf{e}=\textbf{X}^*$ has components which satisfy
\be
{X^A}^* = (-1)^{\textbf{X}A} X^A.
\ee
As a simple example let us spell this out. Using the fact that the degree of $X^A$ with respect to the $\mathbb{Z}_2$ grading is given by $(-1)^{\textbf{X}+A}$, and \eqref{standard} in the final step,
\begin{align}
X^A{}_A\textbf{e} &= (X^A{}_A\textbf{e})^* = {}_A\textbf{e}^*{X^A}^*
=(-1)^{A(\textbf{X}+A)} {X^A}^*{}_A\textbf{e}^* 
= (-1)^{A\textbf{X}} {X^A}^*{}_A\textbf{e}\,.
\end{align}


Normally, the index is enough to define the transformation when we change coordinates. For example for a transformation purely in the bosonic sector:
\be
X^\mu \mapsto {X'}^\mu = X^\nu K^\mu_{\:\:\nu} = X^\nu\frac{\partial x^\mu}{\partial x^\nu}.
\ee
However for supermanifolds, we need to distinguish whether the Jacobian matrix $K$ acts from the left or the right since we are dealing with non-commuting fields. Suppose we change  basis:
\be
{}_A\textbf{e}={}_AK^B{}_B\bar{\textbf{e}}\,.
\ee
Since $\textbf{X}$ has an independent meaning, it must be left unchanged. We are thus led to define
\be
\bar{X}^A = X^B{}_{B\!}K^A\,.
\ee
If  $\{{}_A\textbf{e}\}$ and $\{{}_A\bar{\textbf{e}}\}$ are both standard bases, it follows that if we write $K$ in block form
\be
K=\begin{pmatrix}
A & B\\
C & D
\end{pmatrix}\,,
\ee
the entries of $A$ and $D$ are all c-numbers, and the entries of $B$ and $C$ are all a-numbers. Thus the degree of ${}_AK^B$ is $(-1)^{A+B}$.

With indices defined as both prefixes and suffixes, it begs the question as to what is meant by ${}^A\!X$. We note that:
\be
\bar{X}^A = X^B{}_{B\!}K^A = (-1)^{(\textbf{X}+B)(A+B)}{}_{B\!}K^AX^B = (-1)^{\textbf{X}A}(-1)^{B(A+B)}{}_{B\!}K^A(-1)^{\textbf{X}B}X^B\,,
\ee
and thus we are led to define
\be
\label{upshift}
{}^{A\!}X = (-1)^{\textbf{X}A}X^A\quad\text{and}\quad {}^{A\!}K^{\sim}_{\:\:B} = (-1)^{B(A+B)}{}_{B\!}K^A\,,\quad\hbox{so that}\quad
{}^{A\!}\bar{X} = {}^AK^{\sim}_{\:\:B}\,{}^B\!X\,,
\ee
where we have defined the \emph{supertranspose} of $K$. In order to have $X^A{}_A\textbf{e}=\textbf{e}_A{}^{A\!}X$, we then have
\be
\textbf{e}_A=(-1)^A{}_A\textbf{e}
\ee
as a definition for basis vectors with index on the left. The supertranspose can also be defined for other index placements, and these are:
\be
{}_A{L^{\sim}}^B = (-1)^{A(A+B)}\,{}^B\!L_A,\quad {}_AM^\sim_{\:\:B} = (-1)^{A+B+AB}{}_BM_A,\quad {}^A{N^\sim}^B = (-1)^{AB} \,{}^{B\!}N^A\,.
\ee
With these definitions $K^{\sim\sim}=K$, while a \emph{supersymmetric matrix} is one which satisfies $K^{\sim}=K$.


Let $\{\textbf{e}^A\}$ be the dual basis to $\{{}_A\textbf{e}\}$. Then they act as a basis for forms. 
We write $\boldsymbol{\omega}=\textbf{e}^A{}_A\omega$ and define $\boldsymbol{\omega}(\textbf{X})=X^A{}_A\omega$, where ${}_A\omega$ has the expected degree $(-1)^{\boldsymbol{\omega}+A}$. If we want to have $X^A{}_A\omega=(-1)^{\boldsymbol{\omega}\textbf{X}}\omega_A{}^{A\!}X$ then we must define
\be
\label{downshift}
\omega_A=(-1)^{A(\boldsymbol{\omega}+A)}{}_A\omega\,.
\ee
Note from \eqref{upshift}, the difference in index-shifting conventions between up and down indices. This behaviour carries over to tensors.
For c-type matrices (\ie those of the form of the coordinate transformations) we define
\be
K_A^{\:B} = (-1)^A{}_AK^B,\quad L^A_{\:B}={}^{A\!}L_B,\quad M_{AB} = (-1)^A{}_AM_B,\quad N^{AB} = {}^{A\!}N^B.
\ee
Note that we are only able to move the leftmost right index to the left, and the rightmost left index to the right. This generalises to c-type tensors so that shifting an upper index can be done for free, whereas shifting a lower index comes with a $(-1)^A$.
With this convention we have, for example
\be
K^\sim_{\:\:AB} = (-1)^{AB} K_{BA}
\ee
and all matrices with both indices on the right have the same $(-1)^{AB}$ behaviour under supertransposition, meaning that for a supersymmetric matrix we have
\be
\label{supersymmetricMatrix}
S_{AB} = (-1)^{AB}S_{BA}
\ee
as one might naively expect.

Note that the different index-shifting conventions (\ref{upshift}, \ref{downshift}) require care. For example, 
\be
{}_A\delta^B,\quad {}^A\delta_B,\quad \delta^A_{\:\:B}
\ee
all represent the Kronecker delta, but
$
\delta_{\!A}^{\:B}= (-1)^A{}_A\delta^B
$
does not.
For a matrix with index positions ${}_AK^B$, we define the supertrace as
\be
\label{strone}
\str K =(-1)^A{}_AK^B=K_A^{\:A}\,,
\ee
and similarly for ${}^A\!L_B$:
\be
\label{strtwo}
\str L =(-1)^A\;{}^{A\!}L_A = (-1)^A\, L^A_{\:A}\,.
\ee
Contrasting  (\ref{strone},\ref{strtwo}), we see again that with indices on the right, the ``natural" index placement leads to different behaviour.
One can define the superdeterminant by working with \cite{\DeWitt}
\be
\label{sdet}
\delta\ln\sdet M = \str(M^{-1}\delta M)\,,
\ee
with the condition that $\sdet \mathbb{I}=1$\,, in analogy with the determinant. The result is the Berezinian. In particular 
\be 
\label{sdetBlock}
\sdet \begin{pmatrix}
A & 0\\
0 & B
\end{pmatrix} = \frac{\det A}{\det B}\,.
\ee
As usual, we define vector fields through their action on functions:
\be
\label{leftderiv}
X(f)=X^A\frac{\overrightarrow{\partial}}{\partial x^A}f = X^A{}_{A,}f
\ee
with the obvious notation that indicates that the derivative acts from the left.
Again care is required since the usual notation $f_{,A}$ now means something slightly different. Indeed, we have
\be
\label{rightderiv}
f_{,A}=f\frac{\overleftarrow{\partial}}{\partial x^A}=(-1)^{A(f+1)}{}_{A,}f
\ee
and other similar rules. 


We wish to work in a Riemannian supermanifold and therefore have to define a metric. 
This is a real, c-type, non-degenerate supersymmetric (0,2) tensor $g$.
It defines a natural inner product
\be
\label{innerprod}
g(\textbf{X},\textbf{Y})=\textbf{X}\cu\cdot\textbf{Y}=X^A{}_Ag_B{}^BY = (-1)^{\textbf{X}\textbf{Y}}g(\textbf{Y},\textbf{X})\,,
\ee
from which we get the transposition rules for $g$. They are none other than those of a supersymmetric matrix. In particular if both indices are shifted to the right, then $g$ satisfies \eqref{supersymmetricMatrix}:
\be
\label{gsymm}
g_{AB}=(-1)^{AB}g_{BA}\,.
\ee
We also have the inverse metric ${}^{A\!}g^B=g^{AB}$, again supersymmetric, and which is defined by
\be
{}^{A\!}g^B{}_Bg_C={}^A\delta_C,\quad {}_Ag_B\,{}^{B\!}g^C={}_A\delta^C
\ee
We can use the metric and its inverse to raise and lower indices on vector and tensor fields, however taking care to use only ``natural" contractions:
\be
\label{raiselower}
X_A=X^B{}_Bg_A,\quad X^A=X_B\,{}^Bg^A,\quad {}_AX={}_Ag_B{}^BX,\quad {}^AX={}^Ag^B\,{}_BX\,,
\ee
so that right indices are raised/lowered with the first index on the metric, and left indices are raised with the second index. 


The Riemannian connection coefficients are then 
\be
\label{connection}
\Gamma^A_{\:\:BC}=\frac{(-1)^{D}}{2}g^{AD}\left(g_{DB,C}+(-1)^{BC}g_{DC,B}-(-1)^{D(B+C)}g_{BC,D}\right)\,.
\ee 
In terms of these, the Riemann tensor is given by
\be
\label{Riemann}
R^A_{\:\:BCD}=-\Gamma^{A}_{\:\:BC,D}+(-1)^{CD}\Gamma^A_{\:\:BD,C}+(-1)^{C(E+B)}\Gamma^A_{\:\:EC}\Gamma^E_{\:\:BD}-(-1)^{D(E+B+C)}\Gamma^A_{\:\:ED}\Gamma^E_{\:\:BC}\,,
\ee
and the Ricci tensor and scalar by
\be
\label{Ricci}
R_{AB}=(-1)^{C(A+1)}R^C_{\:\:ACB}\,,\qquad
R=R_{AB}\,g^{BA}\,.
\ee
Finally in order to compute the action of diffeomorphisms we need to the following formulae for Lie derivatives on supermanifolds:
\begin{align}
\label{Lief}
\mathcal{L}_\xi f &=\xi f\\
\label{LieX}
\mathcal{L}_\xi X &= [\xi,X]\\
\label{LieT}
\mathcal{L}_\xi(T(X,Y)) &= (\mathcal{L}_\xi T)(X,Y) + (-1)^{\xi T}T(\mathcal{L}_\xi X,Y) + (-1)^{\xi(T+X)}T(X,\mathcal{L}_\xi Y)
\end{align}
where $\xi, X, Y$ are vector fields,  $f$ is a function and $T$ is a rank-(0,2) tensor on the supermanifold.

\section{Parisi-Sourlas supergravity}
\label{sec:setup}

Recall from sec. \ref{sec:intro}, that the aim is to formulate a spontaneously broken Parisi-Sourlas supergravity as the regularisation structure for a manifestly diffeomorphism-invariant renormalization group equation. We assume that this can be built from the super-Einstein-Hilbert action:
\be
\label{action}
S=-2\int\!\! d^D\theta\, d^D\!x\,\sqrt{g} R/\kappa^2\,,
\ee
where $\kappa=\sqrt{32\pi G}$ is the natural coupling constant, $G$ being Newton's gravitational constant. The factor of $-2$ is the correct factor for the Einstein-Hilbert action in Euclidean signature. It is not so clear that it is the correct factor for a Parisi-Sourlas supergravity action, as we will discuss in sec. \ref{sec:bose}. In order for the wrong-statistics fields to have interactions that mimic as closely as possible the original graviton interactions, we set the torsion to vanish and thus the connection is given by the Riemannian one: \eqref{connection}.

In the current paper we set the cosmological constant term to zero. It may however play a crucial r\^ole as we also point out in sec. \ref{sec:discussion}. \TRM{We take the base manifold to be flat $\mathbb{R}^D$ and discard boundary contributions, and assume a trivial bundle in the fermionic directions. This will be required for a Wilsonian RG analysis, for example fixed points, since the manifold must remain invariant under Kadanoff blocking \cite{Morris:2018mhd}, but this is also the obvious choice for determining the propagating degrees of freedom, as we do in the remainder of the paper.}

Recalling \eqref{innerprod} and  (\ref{upshift},\ref{downshift}), we can now be precise about the formula \eqref{rough}:
\be 
ds^2 = dx^A{}_Ag_B\,{}^B\!dx = dx^A{}_Ag_B\, dx^B = (-1)^A dx^A\, g_{AB}\, dx^B\,.
\ee
Since $g_{AB}$ is a supersymmetric matrix, \ie satisfies \eqref{gsymm}, if we define
\be
g_{AB}=\begin{pmatrix}
g_{\mu\nu} & g_{\mu b}\\
g_{\nu a} & g_{ab}
\end{pmatrix}\,,
\ee
then  $g_{\mu\nu}=g_{\nu\mu}$, $g_{\mu a}=g_{a\mu}$ and $g_{ab}=-g_{ba}$, as already assumed in the Introduction. 

\subsection{Kinetic terms around flat background}

To find out what are the propagating degrees of freedom, we expand around the background metric \eqref{flatg} to isolate the kinetic terms for fluctuations, writing to first order
\be
\label{metricfirstorder}
g_{AB}=\bar{\delta}_{AB}+\kappa\, h_{AB}\,.
\ee
Since $g_{AB}=\bar{\delta}_{AB}$ is trivially a solution to the vacuum super-Einstein equations, the above is sufficient to get the $O(\kappa^0)$ part of \eqref{action}, \ie the bilinear terms for $h_{AB}$. We put aside the factor of $-2$ in \eqref{action}, splitting the Lagrangian density (up to surface terms) as
\be 
\label{sumL}
\mathcal{L}=\sqrt{g}R/\kappa^2 = \mathcal{L}_{bb}+\mathcal{L}_{bm}+\mathcal{L}_{bf}+\mathcal{L}_{mm}+\mathcal{L}_{mf}+\mathcal{L}_{ff}+O(\kappa)\,,
\ee
naming the parts 
according to whether they involve the `bosonic'  fluctuation indices $h_{\mu\nu}$, `mixed' fluctuation indices $h_{\mu a}$, or `fermionic' fluctuation indices $h_{ab}$ \TRM{(labelled above as $b$, $m$, and $f$ respectively. Each of the} matrix components themselves will have component fields \eqref{componentfields} which are both fermionic and bosonic.)

To raise super-indices, we use the background metric, $\bar{\delta}^{AB}$, which recall is the matrix inverse of ${}_A\bar{\delta}_{B}$. In this case it is also consistent to raise bosonic indices with $\delta^{\mu\nu}$ and fermionic indices with $\epsilon^{ab}$ (the matrix inverse of ${}_a\epsilon_b=-\epsilon_{ab}$), taking the conventions on raising indices as detailed in \eqref{raiselower}. From these one can verify that the inverse metric for ${}_A g_B$, is indeed
\be
\label{metricinversefirstorder}
g^{AB} = \bar{\delta}^{AB} -\kappa\,h^{AB}\,,
\ee
just as it is for the purely bosonic case, and where again we have used \eqref{upshift} to collect indices on the right. In \eqref{action}, $g$ is the super-determinant (or Berezinian).
By \eqref{deteps} and \eqref{sdetBlock}, $\sdet\,\bar{\delta}=1$. Then using \eqref{sdet}, we have
\be
\label{cc}
\sqrt{g} = 1 +\frac{\kappa}{2}\, \str({}^A\bar{\delta}^B{}_Bh_C) = 
1+\frac{\kappa}{2}(-1)^A \,h^A_{\:\:A}=1+\frac{\kappa}{2}\left(h^\mu_{\:\:\mu}-h^a_{\:\:a}\right)\,.
\ee

Unpacking \eqref{connection}, we get six connection coefficients $\Gamma^\mu_{\:\:\nu a}=\Gamma^\mu_{\:\:a\nu}$ and $\Gamma^a_{\:\:\nu b}=\Gamma^a_{\:\: b\nu}$, and each of these themselves contain six terms. These in turn are substituted into \eqref{Riemann} and \eqref{Ricci} to give, before collection, approximately a hundred terms. Note that the right-derivatives $g_{AB,C}=(-1)^{C(A+B+1)}\partial_C g_{AB}$, \cf \eqref{rightderiv}, and similarly $\Gamma^A_{\:\:BC,D}=(-1)^{D(A+B+C+1)}\partial_D \Gamma^A_{\:\:BC}$. Although we write here $\partial_A$ with an index on the right in the usual way, this is really a left-index \cf \eqref{leftderiv}. Thus for example,
\be
\partial_a\partial^a = -\partial^a\partial_a\,,\quad\hbox{as with}\quad h^a_{\:\:a} = -h_a^{\:\:a}\,,\quad\hbox{but}\quad \partial_a V^a = \partial^a V_a\,.
\ee
The final result is 
\beal
\label{bbff}
\mathcal{L}_{bb} &=\frac{1}{4}\partial_\rho h^\mu_{\:\:\mu}\partial_\rho h^\nu_{\:\:\nu} + \frac{1}{2}h^\rho_{\:\:\rho}\partial_\mu\partial_\nu h^{\mu\nu} -\frac{1}{4}\partial_\rho h_{\mu\nu}\partial^\rho h^{\mu\nu} + \frac{1}{2}\partial^\nu h_{\mu\nu}\partial_\rho h^{\mu\rho}\nn\\ 
&\phantom{=\frac{1}{4}\partial_\rho h^\mu_{\:\:\mu}\partial_\rho h^\nu_{\:\:\nu} + \frac{1}{2}h^\rho_{\:\:\rho}\partial_\mu\partial_\nu h^{\mu\nu} }\ 
- \frac{1}{4}\partial_a h^\mu_{\:\:\mu}\partial^a h^\nu_{\:\:\nu} + \frac{1}{4}\partial_a h_{\mu\nu}\partial^a h^{\mu\nu}\,,
\nn\\
\mathcal{L}_{bm} &=-h^\mu_{\:\:\mu}\partial_\nu\partial_a h^{\nu a} - \partial^\nu h_{\mu\nu}\partial_a h^{\mu a}\,,\nn\\
\mathcal{L}_{bf} &=\frac{1}{2}\partial_\rho\partial^\rho h^\mu_{\:\:\mu} h^a_{\:\:a} + \frac{1}{2}h^\mu_{\:\:\mu}\partial^b\partial_b h^a_{\:\:a} - \frac{1}{2} h^a_{\:\:a}\partial_\mu\partial_\nu h^{\mu\nu} - \frac{1}{2}h^\mu_{\:\:\mu}\partial_a\partial_b h^{ab}\,,\\
\mathcal{L}_{mm} &= -\frac{1}{2}\partial_\nu h_{\mu a}\partial^\nu h^{\mu a} -\frac{1}{2}\partial_\mu h^{\mu a}\partial^\nu h_{\nu a} -\frac{1}{2}\partial_b h_{\mu a}\partial^b h^{\mu a} + \frac{1}{2}\partial^a h_{\mu a}\partial_b h^{\mu b}\,,\nn\\
\mathcal{L}_{mf} &= h^a_{\:\:a}\partial_\mu\partial_b h^{\mu b} + \partial^\mu h_{\mu a}\partial_b h^{ab}\,,\nn\\
\mathcal{L}_{ff} &= \frac{1}{4}\partial_\mu h^a_{\:\:a}\partial^\mu h^b_{\:\:b}-\frac{1}{4}\partial_c h^a_{\:\:a}\partial^c h^b_{\:\:b} + \frac{1}{2}h^a_{\:\:a}\partial_c\partial_d h^{cd}+\frac{1}{4}\partial_\mu h_{ab}\partial^\mu h^{ab}\nn\\ 
&\phantom{= \frac{1}{4}\partial_\mu h^a_{\:\:a}\partial^\mu h^b_{\:\:b}-\frac{1}{4}\partial_c h^a_{\:\:a}\partial^c h^b_{\:\:b}}\ 
-\frac{1}{4}\partial_c h_{ab}\partial^c h^{ab} + \frac{1}{2}\partial^b h_{ab}\partial_c h^{ac}\,.\nn
\eeal
Note that if we delete the terms with fermionic indices, we are left with the \TRM{top line above. Up to the discarded factor of $-2$ these are the standard graviton kinetic terms, \ie identical in form to the Fierz-Pauli action. For later purposes we write the latter as
\be 
\label{FP}
\mathcal{L}_{FP} = -\frac12\, h_{\alpha\beta} \,\Box^{\alpha\beta,\mu\nu}\, h_{\mu\nu}\,,
\ee
where $\Box^{\alpha\beta,\mu\nu} = \Box\, \delta^{\alpha (\mu}\delta^{\nu) \beta} +\cdots$ is the Fierz-Pauli differential operator.  Of course this form follows from the choice of Einstein-Hilbert form \eqref{action} of our super-action. However as usual it is also fixed uniquely by invariance under linearised diffeomorphisms, or rather here the linearised super-diffeomorphisms carrying purely bosonic indices (see below). The Fierz-Pauli form \eqref{FP} then yields a number of Fierz-Pauli actions for component fields, as we  explain at the end of sec. \ref{sec:decomposition}.}

\subsection{Linearised super-diffeomorphisms}

The above action for free fields is invariant under linearised super-diffeomorphisms applied to \eqref{metricfirstorder}:
\be
\label{diffh}
{}_Ah_B\mapsto {}_Ah_B + {}_A(\mathcal{L}_\xi\bar{\delta})_B\,,
\ee
where the action of the Lie derivative is given by \eqref{LieT}. We now unpack this definition to get the gauge transformations for the fluctuation fields in \eqref{bbff}, and verify that the resulting action \eqref{sumL} is indeed invariant. In the remainder of the paper, the gauge transformations are then used to isolate the true propagating degrees of freedom.

The expressions simplify on noting that the metric is a real $c$-type tensor, and also that for a general supermanifold $M$, the Lie algebra of Diff$(M)$ is generated by objects of the form $\mathcal{L}_\xi$, with $\xi$ a $c$-type vector \cite{\DeWitt}. From \eqref{LieX} we have
\be 
(\mathcal{L}_\xi X)^A = \xi^B{}_{B,}X^A - X^B{}_{B,}\xi^A\qquad\text{and}\qquad 
{}^A(\mathcal{L}_\xi X) = -{}^A\xi_{,B}{}^BX + {}^AX_{,B}{}^B\xi\,,
\ee
where the second equation follows from the first on using index shifting rules (\ref{upshift},\ref{downshift}), or by swopping left for right derivatives. Using also \eqref{Lief} for the LHS of \eqref{LieT}, the Lie derivative of the $(0,2)$ tensor becomes:
\besp 
\xi^C\,{}_{C,}X^{\!A}\,{}_AT_B\,{}^BY + X^{\!A}\,\xi^C\,{}_{C,A}T_B\,{}^BY + X^A\,{}_AT_B\,\xi^C\,{}_{C,}^{\:\:B}Y 
=\\ X^A\,{}_A(\mathcal{L}_\xi T)_B\,{}^BY + (\xi^C\,{}_{C,}X^A-X^C\,{}_{C,}\xi^A)\,{}_AT_B\,{}^BY + X^A\,{}_AT_B(-{}^B\xi_{,C}\,{}^CY + {}^BY_{,C}\,{}^C\xi)\,.
\eesp
The terms where $X$ is differentiated can be seen to cancel. The same for $Y$ after some manipulation. Thus since $X$ and $Y$ are arbitrary supervectors, we find:
\be
\label{LieTdowndown}
{}_A(\mathcal{L}_\xi T)_B = \xi^C{}_{C,A}T_B + {}_{A,}\xi^C{}_CT_B + {}_AT_C\,{}^C\xi_{,B}\,.
\ee
Therefore from \eqref{diffh} we have
\be
\label{diffhdowndown}
{}_A(\delta_\xi h)_{B} = {}_{A,}\xi_B + {}_A\xi_{,B}\,,
\ee
and thus specialising the indices we have:
\be
\label{specialdowndown}
(\delta_\xi h)_{\mu\nu} = \partial_\mu \xi_\nu + \partial_\nu \xi_\mu\,,\quad
(\delta_\xi h)_{\mu a} = \partial_\mu \xi_a - \partial_a \xi_\mu = (\delta_\xi h)_{a \mu}\,,\quad
(\delta_\xi h)_{ab} = -\partial_a\xi_b + \partial_b\xi_a\,,
\ee
the first equation of course being the usual formula for linearised bosonic diffeomorphisms. \TRM{As we already commented, this first equation, together with the requirement that the action has two derivatives, is sufficient to guarantee (up to proportionality) the Fierz-Pauli form \eqref{FP} for that part of the action dependent on purely bosonic indices.}

Since at the linearised level, \eqref{diffhdowndown} is again a tensor, using $\bar\delta$ and the rules \eqref{raiselower}, we also get
\beal
\label{diffupdown}
(\delta_\xi h)^A_{\:\:B}&= \partial^A\xi_B+(-1)^{B(A+1)}\partial_B\xi^A\,,\\
\label{diffupup}
(\delta_\xi h)^{AB}&= \partial^A \xi^B+(-1)^{AB}\partial^B\xi^A \,,
\eeal
and thus
\be
\label{specialupup}
(\delta_\xi h)^{\mu\nu} = \partial^\mu\xi^\nu + \partial^\nu\xi^\mu\,,\quad
(\delta_\xi h)^{\mu a} = \partial^\mu\xi^a + \partial^a\xi^\mu = (\delta_\xi h)^{a\mu}\,,\quad
(\delta_\xi h)^{ab} = \partial^a\xi^b - \partial^b\xi^a\,,
\ee
while summing over bosonic (fermionic) indices separately, gives:
\be
\label{specialupdown}
(\delta_\xi h)^\mu_{\:\:\mu} = 2\partial_\mu\xi^\mu,\qquad (\delta_\xi h)^a_{\:\:a} = 2\partial_a\xi^a\,.
\ee
Applying \eqref{specialdowndown}, \eqref{specialupup}, \eqref{specialupdown} to \eqref{bbff} gives (up to integration by parts):
\begin{align}
\delta_\xi\mathcal{L}_{bb} &= -\partial_a h^\mu_{\:\:\mu}\partial^a\partial^\nu\xi_\nu + \partial_a h^{\mu\nu}\partial^a\partial^\mu\xi^\nu\,,\nn\\
\delta_\xi\mathcal{L}_{bm} &= -\partial_\mu\xi^\mu\partial_\nu\partial_a h^{\nu a} - h^\mu_{\:\:\mu}\partial_\nu\partial_a(\partial^\nu\xi^a + \partial^a\xi^\nu) - \partial^\nu\partial_\nu\xi_\mu\partial_a h^{\mu a} 
- \partial^\nu h_{\mu\nu}\partial_a(\partial^\mu\xi^a+\partial^a\xi^\mu)\,,\nn\\
\delta_\xi\mathcal{L}_{bf} &= \partial_\rho\partial^\rho h^\mu_{\:\:\mu}\partial_a\xi^a + \partial_\mu\xi^\mu\partial^b\partial_b h^a_{\:\:a} - \partial_a\xi^a\partial_\mu\partial_\nu h^{\mu\nu} - \partial_\mu\xi^\mu\partial_a\partial_b h^{ab}\,,\nn\\
\delta_\xi\mathcal{L}_{mm} &= -\partial_\nu h_{\mu a}\partial^\nu(\partial^\mu\xi^a + \partial^a\xi^\mu) - \partial_\nu h^{\nu a}\partial^\mu(\partial_\mu \xi_a - \partial_a \xi_\mu)\label{bits}\\ 
&\phantom{-\partial_\nu h_{\mu a}\partial^\nu(\partial^\mu\xi^a}
- \partial_b h_{\mu a}\partial^b(\partial^\mu \xi^a + \partial^a \xi^\mu) + \partial^a h_{\mu a}\partial_b(\partial^\mu\xi^b + \partial^b\xi^\mu)\,,\nn\\
\delta_\xi\mathcal{L}_{mf} &= \partial_a\xi^a\partial_\mu\partial_b h^{\mu b} + h^a_{\:\:a}\partial_\mu\partial_b(\partial^\mu \xi^b + \partial^b\xi^\mu) + \partial^\mu(\partial_\mu\xi_a + \partial_a\xi_\mu)\partial_b h^{ab} - \partial^\mu h_{\mu a}\partial_b\partial^b\xi^a\,,\nn\\
\delta_\xi\mathcal{L}_{ff} &= \partial_\mu h^a_{\:\:a}\partial^\mu \partial_b\xi^a + \partial_\mu h_{ab}\partial^\mu\partial^a\xi^b\,.\nn
\end{align}
Adding all of these together and comparing similar terms, confirms that $\delta_\xi\mathcal{L}=0$ up to surface terms, \ie that the linearised action is gauge-invariant.

Finally, we remark that formula \eqref{diffupup} can alternatively be derived from the Lie derivative of the inverse metric ${}^A(\mathcal{L}_\xi g)^B$, using the fact that its expansion, \eqref{metricinversefirstorder}, then implies ${}^A(\delta_\xi h)^B = -{}^A(\mathcal{L}_\xi\bar{\delta})^B$. For this we need also that for a (2,0)-tensor:
\be
{}^A(\mathcal{L}_\xi T)^B = \xi^C{}_{C,}^{\:\:A}T^B - {}^A\xi_{,C}{}^CT^B - {}^AT^C{}_{C,}\xi^B\,.
\ee
This latter expression follows in a similar way to \eqref{LieTdowndown}, namely via
\be
\mathcal{L}_\xi (T(\omega,\chi)) = (\mathcal{L}_\xi T)(\omega,\chi) + T(\mathcal{L}_\xi \omega,\chi) + T(\omega,\mathcal{L}_\xi \chi)\,,
\ee
which holds for an arbitrary c-type (2,0)-tensor $T$ and one-forms $\omega$ and $\chi$, and on using the fact that the Lie derivative of a one-form is given by
\begin{align}
{}_A(\mathcal{L}_\xi\omega) &= {}_{A,}\xi^B{}_B\omega + \xi^B\,{}_{B,A}\omega\,,\\
(\mathcal{L}_\xi\omega)_A &= \xi^B{}_{B,}\omega_A + \omega_B\,{}^B\xi_{,A}
\end{align}
(as follows from considering $\omega(X)$ for arbitrary $X$). The result is again \eqref{diffupup}.

\section{Field Decomposition}
\label{sec:decomposition}

So far we have been working in general dimension $D$. At this point it becomes convenient to specialise to the dimension of interest, $D\cu=4$, which we will do from now on.
Since $h$ is a field on the supermanifold, it can be Taylor expanded in the $\theta$ coordinates as follows:
\be
\label{componentfields}
h(x,\theta) = h(x) + M\theta^a h_{|a}(x) + M^2\theta^a\theta^b h_{|ab}(x) + M^3\theta^a\theta^b\theta^c h_{|abc}(x) + M^4\theta^a\theta^b\theta^c\theta^d h_{|abcd}(x)
\ee
(we will omit the arguments where there is no confusion).  Since we have set $D\cu=4$, and $\theta^a$ is Grassmann, the expansion stops at $\theta^4$. The vertical bar is there to distinguish between this superfield expansion and the spacetime indices on $h_{AB}$, which latter we temporarily suppress. 

To keep numbers simple in the following, the component fields absorb a factor of $1/n!$ compared to the Taylor expansion coefficients:
\be 
M^n\, h_{|a_1\cdots a_n}(x) = \frac1{n!}\, \partial_{a_n}\cdots\partial_{a_1} h(x,\theta)\ \Big|_{\theta=0}\,.
\ee
We  assign mass dimension $[\theta^a]=-1$ to the fermionic coordinates so that the supercoordinates $x^A$ in \eqref{supercoords}, have definite mass dimension. As discussed in the Introduction, the component fields $h_{|a_1\cdots a_n}$ are destined to become part of the regulating structure for the graviton. It is convenient then to keep them all the same dimension as $[h(x,\theta)]=[h_{\mu\nu}]=1$. This is why we introduce an arbitrary mass scale  $M$.

On integrating over $d^4\theta$, the only non-vanishing terms are those with exactly 4 powers of $\theta$. These are then proportional to $\epsilon^{abcd}$, the Levi-Civita symbol in 4 dimensions. The proportionality constant is our choice in defining the measure. We thus set 
\be
\label{measure}
\int\!\!\! d^4\theta \ \theta^a\theta^b\theta^c\theta^d = M^{-4}\,\epsilon^{abcd}\,,
\ee
where the factor $M^{-4}$ together with \eqref{componentfields} ensures that terms with two space-time derivatives (\ie bosonic $\partial_\mu$) come out correct dimensionally. 

The formulae also come out neater, if we utilise a Hodge dual in the $\theta$ space:\footnote{\TRM{Notice that Hodge duality here plays a purely algebraic r\^ole. There are no topological implications unlike its use for forms in standard supergravity. (See also comments at the end of the paper.)}}
\be
\label{hodge}
*\!h = \epsilon^{abcd} h_{|abcd}\,,\quad
*h^{|a} = \epsilon^{abcd} h_{|bcd}\,,\quad
*h^{|ab} = \frac{1}{2}\epsilon^{abcd} h_{|cd}\,.
\ee
For completeness we further define
\be 
*\!h^{|abc} = \frac{1}{6}\epsilon^{abcd}h_{|d}\,,\quad
*h^{|abcd} = \frac{1}{24}\epsilon^{abcd}h\,,
\ee
and for neatness define $*\partial_\mu h =\partial_\mu\!\!*\! h$. We
define the Hodge dual of the lower index expressions in the same way, using the Levi-Civita symbol written as $\epsilon_{abcd}$. Thus we have
\be
\label{hodgesquared}
*\!(*h)_{|{a_1}\dots{a_n}} = (-1)^{n}h_{|{a_1}\dots{a_n}}\,,
\ee
as expected for an even number of dimensions. We will also make use of the following standard formula:
\be
\epsilon^{i_1\dots i_k i_{k+1}\dots i_4}\epsilon_{i_1\dots i_k j_{k+1}\dots j_4} = k!\,\delta^{i_{k+1}\dots i_4}_{j_{k+1}\dots j_4}
\ee
where the generalised Kronecker $\delta$ is the sum over all products of Kronecker $\delta^{i_n}_{j_m}$ including the sign of the permutation required to get from the upper to the lower indices. Then for any two metric components $h$ and $h'$, we have
\beal
\int\!\! d^4\theta \,  \partial_\mu h\partial_\nu h' &= \partial_\mu h\cu*\partial_\nu  h' - (-1)^{h}\partial_\mu h_{|a}\cu*\partial_\nu {h'}^{|a} + 2\partial_\mu h_{|ab}\cu*\partial_\nu {h'}^{|ab} + (-1)^{h}\cu* \partial_\mu h^{|a}\partial_\nu h'_{|a} + *\partial_\mu h\partial_\nu h',\nn\\
\int\!\! d^4\theta\, \partial_\mu h\partial_a h' &= -(-1)^{h}M\partial_\mu h_{|a} \cu*h' - 2M\partial_\mu h_{|ab} {*h'}^{|b} +2(-1)^{h}M\cu*\partial_\mu h^{|b} h'_{|ab} 
+ M\cu*\partial_\mu h \,h'_{|a}\,,\\
\int\!\! d^4\theta\, \partial_a h\partial_\mu h' &= M h_{|a}\cu*\partial_\mu h' + 2(-1)^hM h_{|ab}\cu*\partial_\mu {h'}^{|b} - 2M\cu*h^{|b}\partial_\mu h'_{|ab} 
- (-1)^hM\cu*h\partial_\mu h'_{|a}\,,\nn\\
\int\!\! d^4\theta\, \partial_a h\partial_b h' &= 2(-1)^hM^2 h_{|ab}\cu*h' - M^2\epsilon_{abcd}\cu*h^{|c}{*h'}^{|d} + 2(-1)^hM^2 \cu*h\, h'_{|ab}\nn
\eeal
(the third equation also following from the second by symmetry). Thus, expanding the part of the action with bosonic metric components, we get 
\besp
\label{bb}
\mathcal{L}_{bb} = 2\partial_\rho\varphi\,\cu*\partial^\rho\!\varphi -2\partial_\mu\varphi_{|a}\partial^\mu\varphi^{|a} + 2\partial_\rho \varphi_{|ab}\cu*\partial^\rho\! \varphi^{|ab} 
 +\varphi\,\cu*\partial_\mu\partial_\nu h^{\mu\nu} - \varphi_{|a}\partial_\mu\partial_\nu *h^{\mu\nu|a} \\
+ 2\varphi_{|ab}\cu*\partial_\mu\partial_\nu h^{\mu\nu|ab} 
+ *\varphi^{|a}\partial_\mu\partial_\nu h^{\mu\nu}{}_{|a} 
+ *\varphi\,\partial_\mu\partial_\nu h^{\mu\nu} 
-\frac{1}{2}\partial_\rho h_{\mu\nu}\cu*\partial^\rho h_{\mu\nu} +\frac{1}{2}\partial_\rho h_{\mu\nu|\new{a}}\cu*\partial^\rho h^{\mu\nu|\new{a}}\\ 
- \frac{1}{2}\partial_\rho h_{\mu\nu|ab}\cu*\partial^\rho h^{\mu\nu|ab} 
+\partial^\nu h_{\mu\nu}\cu*\partial_\rho h^{\mu\rho} -\partial^\nu h_{\mu\nu|a}\cu*\partial_\rho h^{\mu\rho|a} + \partial^\nu h_{\mu\nu|ab}\cu*\partial_\rho h^{\mu\rho|ab}\\
 -\epsilon^{ab}M^2\left(4\varphi_{|ab} \cu*\varphi - \epsilon_{abcd} \cu*\varphi^{|c}\cu*\varphi^{|d}
 -h_{\mu\nu|ab} \cu*h^{\mu\nu} + \frac{1}{4}\epsilon_{abcd}\cu*h_{\mu\nu}{}^{|c} \cu*h^{\mu\nu|d}\right)\,,
\eesp
where we write  
\be 
\label{phi}
\varphi = \frac{1}{2}h^\mu_{\:\:\mu}\,. 
\ee
We have also taken the opportunity to make the inverse metric component $\epsilon^{ab}$ explicit, where in \eqref{bbff} it was used to raise an index. Similarly we find
\beal
\mathcal{L}_{bm} &= M\bigg(2\partial_\mu\varphi_{|a}\cu*h^{\mu a} - 4\partial_\mu \varphi_{|ab}\cu*h^{\mu a|b} - 4\cu*\partial_\mu \varphi^{|b} h^{\mu a}{}_{|ab} + 2\cu*\partial_\mu\varphi h^{\mu a}{}_{|a}\nn\\
&\qquad\qquad -\partial^\nu h_{\mu\nu|a}\cu*h^{\mu a}
 + 2\partial^\nu h_{\mu\nu|ab}\cu*h^{\mu a|b} + 2\cu*\partial^\nu h_{\mu\nu}{}^{|b} h^{\mu a}{}_{|ab} - *\partial^\nu h_{\mu\nu} h^{\mu a}{}_{|a}\bigg)\,,\nn\\
\mathcal{L}_{bf} &= 2\varphi\cu*\square \chi - 2\varphi_{|a}\cu*\square \chi^{|a} + 4\varphi_{|ab}\cu*\square \chi^{|ab} + 2\cu*\varphi^{|a}\square\chi_{|a} + 2\cu*\varphi\square\chi
+\partial_\mu\chi\cu*\partial_\nu h^{\mu\nu} \\
&\qquad- \partial_\mu \chi_{|a}\cu*\partial_\nu h^{\mu\nu|a} + 2\partial_\mu\chi_{|ab}\cu*\partial_\nu h^{\mu\nu|ab} + *\partial_\mu \chi^{|a}\partial_\nu h^{\mu\nu}{}_{|a} +* \partial_\mu \chi\partial_\nu h^{\mu\nu}\nn\\
&+M^2\left(\epsilon^{ab}\left[ 4\chi_{|ab}\cu*\varphi - 2\epsilon_{abcd}\cu*\chi^{|c}\cu*\varphi^{|d} + 4 \cu*\chi\varphi_{|ab}\right]
 +2\varphi_{|ab}\cu*h^{ab} - \epsilon_{abcd}\cu*\varphi^{|c}\cu*h^{ab|d} + 2\cu*\varphi h^{ab}{}_{|ab}\right)\,,\nn
\eeal
where we have also written 
\be
\label{chi}
\chi = \frac{1}{2}h^a{}_a\,. 
\ee
And finally, 
\begin{align}
\label{mmmfff}
\mathcal{L}_{mm} &= -\partial_\nu h_{\mu a}\cu*\partial^\nu h^{\mu a} - \partial_\nu h_{\mu a|b}\cu*\partial^\nu h^{\mu a|b}- \partial_\nu h_{\mu a|bc}\cu*\partial^\nu h^{\mu a|bc}\nn\\
&\qquad   -\partial_\mu h^{\mu a}\cu*\partial^\nu h_{\nu a} - \partial^\mu h_{\mu a|b}\cu*\partial_\nu h^{\nu a|b} 
- \partial_\mu h^{\mu a}{}_{|bc}\cu*\partial^\nu h_{\nu a}{}^{|bc} 
\nn\\
& -\frac{1}{2}M^2\epsilon^{ab}\left(4\cu*h^{\mu c} h_{\mu c|ab} - \epsilon_{abcd} \cu*h^{\mu e|c}\cu*h_{\mu e}{}^{|d} -4h_\mu{}^a{}_{|ab}\cu*h^{\mu b} + \epsilon_{abcd} \cu*h_\mu{}^{a|c}\cu*h^{\mu b|d}\right)\,,\nn\\
\mathcal{L}_{mf} &= M\bigg(\new{2}\partial_\mu\chi_{|a}\cu*h^{\mu a} +4\partial_\mu \chi_{|ab}\cu*h^{\mu a|b} \new{-4} \cu*\partial_\mu \chi^{|\new{b}}h^{\mu a}{}_{|ab} \new{-2} *\partial_\mu \chi h^{\mu a}{}_{|a}\nn\\
&\quad\quad\quad +\partial^\mu h_{\mu a|b} \cu*h^{ab} -2\partial^\mu h_{\mu a|bc} \cu* h^{ab|c} \new{-} 2\cu*\partial_\mu h_{\mu a}{}^{|c} h^{ab}{}_{|bc} + *\partial^\mu h_{\mu a} h^{ab}{}_{|b}\bigg)\,,\nn\\
\mathcal{L}_{ff} &= 2\partial_\mu\chi\cu*\partial^\mu \chi - 2\partial_\mu \chi_{|a}\cu*\partial^\mu \chi^{|a} + 2\partial_\mu \chi_{|ab}\cu*\partial^\mu \chi^{|ab}
\\
&\quad +\frac{1}{2}\partial_\mu h_{ab}\cu*\partial^\mu h^{ab} - \frac{1}{2}\partial_\mu h_{ab|c}\cu*\partial^\mu h^{ab|c} + \frac{1}{2}\partial_\mu h_{ab|cd}\cu*\partial^\mu h^{ab|cd}
\nn\\
&\quad -M^2\epsilon^{ab}\left(4\chi_{|ab}\cu*\chi - \epsilon_{abcd}\cu*\chi^{|c}\cu*\chi^{|d}
+h_{cd|ab}\cu*h^{cd} - \frac{1}{4}\epsilon_{abcd} \cu*h_{ef}{}^{|c}\cu*h^{ef|d}
\right)
\nn\\
&\quad +M^2\left(2\chi_{|ab}\cu*h^{ab} + 2\cu*\chi h^{ab} - \epsilon_{abcd} \cu*\chi^{|c}\cu*h^{ab|d}
-2h_{a}{}^b{}_{|bc}\cu*h^{ac} + \frac{1}{2}\epsilon_{cdef}\cu*h_a{}^{c|e} \cu*h^{ad|f}
\right)\,.\nn
\end{align}

\TRM{This system can now be recast as graviton fields plus partners, where the latter are to form part of the regulating structure. In particular note that inside $\mathcal{L}_{bb}$ as written in \eqref{bb} are a number of copies of the standard Fierz-Pauli action for graviton fields. They now appear as\footnote{\TRM{There is also $h_{\mu\nu | a}\,\Box^{\alpha\beta,\mu\nu} \cu*h_{\mu\nu}{}^{|a}$. These fields are fermionic so do not correspond to gravitons.}} 
\be 
\label{candidates}
\frac12\, h_{\alpha\beta} \,\Box^{\alpha\beta,\mu\nu}\, \cu*h_{\mu\nu} +\frac14\,\epsilon^{abcd} h_{\alpha\beta | ab} \,\Box^{\alpha\beta,\mu\nu}\, h_{\mu\nu | cd}\,.
\ee
The appearance of the Fierz-Pauli operator \eqref{FP} for any component of the super-field $h_{\mu\nu}$,  is guaranteed by the standard (bosonic) diffeomorphism invariance generated by $\xi_\mu(x)$. Although $\Box^{\alpha\beta,\mu\nu}$ now couples different fields on the left and right, the above action can be diagonalised. The first term then yields a correct sign Fierz-Pauli action for one diagonal component and thus a candidate for the graviton, and a wrong sign Fierz-Pauli action for the other component, which is thus a ghost (as we will see worked out effectively later in \eqref{hpm} and \eqref{threesys}). The second term likewise provides a further six fields, three of which will have the right sign action and thus at this stage are also candidate gravitons, and a further three which are ghosts.

Apparently we should conclude that there are thus four separate graviton fields at the free level. This is the wrong answer. There is just one, or two, depending on some choices of sign. The mistake is to ignore the mixing of these graviton-like fields to the other fields in the theory, as appear in the rest of \eqref{bb} and the other sectors displayed above, and also to ignore the larger local invariances provided by the full super-diffeomorphism invariance. These two effects imply that some of these degrees of freedom propagate in a way that cannot be identified with the graviton, while others are pure gauge.}

\section{Gauge fixing}
\label{sec:gf}

\TRM{In fact, to understand correctly what are the real candidate gravitons and what are their partners, we need first to fix the super-gauge invariance.
To get a clear picture,} we leave till last the standard (bosonic) diffeomorphism invariance generated by $\xi_\mu(x)$, but remove redundancy by using $\xi_a(x,\theta)$ and all the other components of $\xi_\mu(x,\theta)$ to eliminate as many degrees of freedom as we can. 

In analogy with spontaneously broken gauge symmetry, we begin by choosing ``unitary gauge" where possible, \ie proceed by the local algebraic elimination of fields. Finally to fully isolate propagating degrees of freedom in this theory, we will fix to a radiation gauge, first for a remaining supersymmetry, and then for standard diffeomorphisms.

Starting with  \eqref{specialdowndown}, we have 
\beal
\label{habgauge}
&\delta_\xi h_{ab}(x) = 2M\xi_{[a|b]}(x)\,,\quad
\delta_\xi h_{ab|c} = 4M\xi_{[a|b]c}\,,\quad
\delta_\xi h_{ab|cd} = 6M\xi_{[a|b]cd}\,,\quad
\delta_\xi h_{ab|cde} = 8M\xi_{[a|b]cde}\,,\nn\\
&\delta_\xi h_{ab|cdef}(x) = 0\,.
\eeal
In all cases on the first line, the RHS takes the most general form for a function that is antisymmetric in $a$ and $b$. Therefore we can fix $\xi_{[a|b]\cdots}:=\frac12(\xi_{a|b\cdots}-\xi_{b|a\cdots})$ so that all $h_{ab|\dots}=0$, \ie are eliminated, except for  $h_{ab|cdef}$ which normally appears as $*h_{ab}$ and is gauge invariant. 

Next from \eqref{specialdowndown}, we look at
\begin{align}
\label{hmuagauge}
&\delta_\xi h_{\mu a}(x) = \partial_\mu \xi_a(x) - M\xi_{\mu|a}(x)\,,\quad
\delta_\xi h_{\mu a|b} = \partial_\mu \xi_{a|b} - 2M\xi_{\mu|ab
}\,,\quad
\delta_\xi h_{\mu a|bc} = \partial_\mu \xi_{a|bc} - 3M\xi_{\mu|abc}\,,\nn\\
&\delta_\xi h_{\mu a|bcd} = \partial_\mu \xi_{a|bcd} - 4M\xi_{\mu|abcd}\,,\quad
\delta_\xi h_{\mu a|bcde} = \partial_\mu \xi_{a|bcde}\,.
\end{align}
We see that we can fix $\xi_{\mu|a}$ to eliminate $h_{\mu a}$, and we can also fix $\xi_{\mu|ab\cdots}$ to set $h_{\mu[a|b]\cdots}=0$, with the exception of $h_{\mu a|bcde}$, similar to above.
At this point note that there is no tensor  that has symmetry on the first two indices and antisymmetry on the second two. Indeed, in such a case we would have
\be
T_{abc} = T_{bac} = -T_{bca} = -T_{cba} = T_{bca} = -T_{bac} = -T_{abc}
\ee
and thus $T_{abc}=0$. This means that $h_{\mu a | bc\cdots}$ and $\xi_{a|bc\cdots}$ are automatically antisymmetric in $a$ and $b$. Therefore we have actually entirely eliminated  $*h_{\mu a}{}^{|b}$ and $h_{\mu a|bc}$ (equivalently $*h_{\mu a}{}^{|bc}$). For the same reasons, the final gauge transformation in \eqref{hmuagauge} no longer exists, $\xi_{a|bcde}$ having been fixed entirely by setting $h_{ab|cde}=0$ in \eqref{habgauge}. We see therefore that the remaining $*h_{\mu a}$ is now invariant, and the only other remaining field components are $h_{\mu (a|b)} =\frac12 (h_{\mu a|b}+h_{\mu b|a})$, which transform as
\be
\label{hmuabtrans}
\delta_\xi h_{\mu (a|b)} = \partial_\mu \xi_{(a|b)}\,.
\ee
Although no longer active in changing $h_{\mu a}$, we still have a gauge invariance generated by $\xi_a(x)$. But by \eqref{hmuagauge}, any further change $\delta\xi_a =\xi'_a$, must be accompanied by $\delta\xi_{\mu|a}=\xi'_{\mu|a}$, such that
\be 
\label{xip}
\xi'_{\mu|a} = \frac1M\partial_\mu\xi'_a\,,
\ee
in order to maintain $h_{\mu a}(x)=0$.

Evidently from \eqref{bb}, $\mathcal{L}_{bb}$ is unchanged by the above partial gauge fixing, but all other parts of the Lagrangian are profoundly altered. Firstly, we can clearly see that in \eqref{mmmfff}, $\mathcal{L}_{mm}=\mathcal{L}_{ff}=0$, as at least one component in every bilinear now vanishes. Similarly, noting for example that now $h^{\mu a}{}_{|a}=0$ (since the matrix used to raise $a$ is antisymmetric), $\mathcal{L}_{mf}=0$, while
$\mathcal{L}_{bm}$ collapses to 
\be
\label{Lbm}
\mathcal{L}_{bm} = M\left(2\partial_\mu\varphi_{|a}\cu*h^{\mu a} -\partial^\nu h_{\mu\nu|a}\cu*h^{\mu a}\right)\,,
\ee
and, by eliminating all but $*h_{ab}$ or $*\chi$ in \eqref{habgauge}, we are left in $\mathcal{L}_{bf}$ only with:
\be
\mathcal{L}_{bf} = 2\varphi\square \cu*\chi + *\partial_\mu \chi \partial_\nu h^{\mu\nu} + 4M^2\epsilon^{ab} \cu*\chi\varphi_{|ab} + 2M^2\varphi_{|ab}\cu*h^{ab}\,.
\ee
Note that although we still have the mixed fluctuation field and gauge transformation \eqref{hmuabtrans}, we see that the remaining terms \eqref{Lbm} in the free action, do not depend on them. However at the interacting level these degrees of freedom could thus act as Lagrange multipliers, leading to important constraints.

This is as far as we can go purely algebraically. Next we note from \eqref{specialdowndown} and \eqref{xip} that we have the remaining gauge invariance
\be 
\label{xipha}
\delta_{\xi'} h_{\mu\nu|a} = \partial_\mu \xi'_{\nu|a} + \partial_\nu \xi'_{\mu|a} = \frac{2}{M} \partial_\mu\partial_\nu \xi'_a\qquad\implies\qquad \delta_{\xi'} \varphi_{|a} = \frac{\square}{M}\xi'_{a}\,,
\ee
and with a Green's function, we can use $\xi'_a$ to fix the radiation-type gauge $\varphi_{|a}=0$. This reduces both $\mathcal{L}_{bb}$, \cf \eqref{bb}, and $\mathcal{L}_{bm}$ above.
Since we still have $\xi_{\mu}(x)$ gauge invariance, we are now free to choose traceless transverse gauge for $h_{\mu\nu}(x)$, which means in particular that $\varphi=0$, \cf \eqref{phi}. This also removes terms from $\mathcal{L}_{bf}$.

Apart from \eqref{hmuabtrans} which plays no r\^ole at the free level, we have at this point completely fixed the gauge invariance. Combining all remaining terms we can re-express  \eqref{sumL} as $\mathcal{L}=\mathcal{L}_o+\mathcal{L}_e$ where 
\begin{align}
\mathcal{L}_e &= 2\partial_\mu\varphi_{|ab} \cu*\partial^\mu\varphi^{|ab} + 2\varphi_{|ab}\cu*\partial_\mu\partial_\nu  h^{\mu\nu|ab} - \frac{1}{2}\partial_\rho h_{\mu\nu}\cu*\partial^\rho h^{\mu\nu}-\frac{1}{2}\partial_\rho h_{\mu\nu|ab}\cu*\partial^\rho h^{\mu\nu|ab}\\
&+\partial^\nu h_{\mu\nu|ab}\cu*\partial_\rho h^{\mu\rho|ab} -4M^2\epsilon^{ab}\varphi_{|ab}\cu*\varphi+M^2\epsilon^{ab}h_{\mu\nu|ab}\cu*h^{\mu\nu} + 4M^2\epsilon^{ab}\varphi_{|ab}\cu*\chi+2M^2\varphi_{|ab}\cu*h^{ab}\,,\nn\\
\label{LoOrig}
\mathcal{L}_o &= *\varphi^{|a}\partial_\mu\partial_\nu h^{\mu\nu}{}_{|a} +\new{\frac12} \partial_\rho h_{\mu\nu|a}\cu*\partial^\rho h^{\mu\nu|a} - \partial^\nu h_{\mu\nu|a}\cu*\partial_\rho h^{\mu\rho|a} - M\partial^\nu h_{\mu\nu|a}\cu*h^{\mu a}\\ 
&\qquad+ M^2\epsilon^{ab}\epsilon_{abcd}\, \cu*\varphi^{|c}\cu*\varphi^{|d} -\frac{1}{4}M^2\epsilon^{ab}\epsilon_{abcd}\, \cu*h_{\mu\nu}{}^{|c}\cu*h^{\mu\nu|d}\,,\nn
\end{align}
collects together the bosonic field kinetic terms, and fermionic field kinetic terms, respectively.

\section{Propagating degrees of freedom}
\label{sec:props}
\subsection{Bosonic sector}
\label{sec:bose}

With the Lagrangian now in this form, we note that $*h^{ab}$ (and $\cu*\chi$, but this is part of $*h^{ab}$) acts as a Lagrange multiplier, imposing the condition:
\be 
\varphi_{|ab}+\epsilon_{ab}\,\epsilon^{cd}\varphi_{|cd} = 0\,.
\ee
Contracting with $\epsilon^{ab}$ (and using $\epsilon^{ab}\epsilon_{ab}=4$) then establishes that $\varphi_{|ab}=0$,  thus $\mathcal{L}_e$ collapses to:
\be
\label{Le}
\mathcal{L}_e =  - \frac{1}{2}\partial_\rho h_{\mu\nu}\cu*\partial^\rho h^{\mu\nu}-\frac{1}{2}\partial_\rho h_{\mu\nu|ab}\cu*\partial^\rho h^{\mu\nu|ab}
+\partial^\nu h_{\mu\nu|ab}\cu*\partial_\rho h^{\mu\rho|ab} +M^2\epsilon^{ab}h_{\mu\nu|ab}\cu*h^{\mu\nu}\,.
\ee
Note that $h_{\mu\nu|ab}$, and $*h^{\mu\nu|ab}$ are now traceless on their first two indices.

To diagonalise the remaining bosonic kinetic terms in \eqref{Le} it is clearly now helpful to write
\be 
\label{hsplit}
h_{\mu\nu|ab} = \frac12\epsilon_{ab}\, h^\parallel_{\mu\nu}+h^\perp_{\mu\nu|ab}\,,
\ee
where
\be 
\label{hsplitdefs}
\epsilon^{ab}h^\perp_{\mu\nu|ab} = 0\qquad\iff\qquad h^\parallel_{\mu\nu} = \frac12\epsilon^{ab} h_{\mu\nu|ab}\,.
\ee
However we see from \eqref{Le} that this will result in the appearance of $*\epsilon^{ab}$, the Hodge dual of $\epsilon_{ab}$.

In fact this Hodge dual is proportional to the inverse metric $\epsilon^{ab}$:
\be 
\label{epsrelation}
\cu*\epsilon^{ab} = \frac12\epsilon^{abcd}\epsilon_{cd} = s\, \epsilon^{ab}\,.
\ee
Here we recall the definition \eqref{hodge}, and introduce $s$, the Pfaffian of $\epsilon_{ab}$:
\be 
\label{s}
s = \frac18\epsilon^{abcd}\epsilon_{ab}\epsilon_{cd}\,.
\ee
This identity \eqref{epsrelation} actually holds for any $4\cu\times4$ invertible antisymmetric matrix ${}_a\epsilon_b =-\epsilon_{ab}$ and its inverse $\epsilon^{ab}$. It is most easily seen by first rotating to a basis in which $\epsilon_{ab}$ is block-diagonal:
\be 
\epsilon_{ab} = \lambda_1i\sigma_2\oplus \lambda_2i\sigma_2\quad\implies\quad s=\lambda_1\lambda_2\quad \text{and}\quad \epsilon^{ab} =i\sigma_2/\lambda_1\oplus i\sigma_2/\lambda_2\,,
\ee
$i\sigma_2$ being the totally antisymmetric symbol in two dimensions, $\sigma_2$ being the $2^\text{nd}$ Pauli matrix. Finally note that, since the Pfaffian satisfies $\det\epsilon = s^2$ and since we have normalised $\det\epsilon=1$, \cf \eqref{deteps}, we actually have $s=\pm1$.

Taking the Hodge dual of \eqref{hsplit} and using \eqref{epsrelation}, we get
\be 
\label{starhsplit}
\cu*h_{\mu\nu}{}^{|ab} = \frac{s}2\epsilon^{ab}\, h^\parallel_{\mu\nu}+*h^\perp_{\mu\nu}{}^{|ab}\,.
\ee
Contracting with $\epsilon_{ab}$ and using \eqref{epsrelation} and  \eqref{hsplitdefs} 
then establishes that
\be 
\label{starhdefs}
\cu*h_{\mu\nu}{}^{|ab}\epsilon_{ab} = 2s\, h^\parallel_{\mu\nu}\qquad\text{and}\qquad
\cu*h^\perp_{\mu\nu}{}^{|ab}\epsilon_{ab} = 0\,.
\ee
Now substituting \eqref{hsplit} and \eqref{starhsplit} into \eqref{Le}, and using \eqref{hsplitdefs} and \eqref{starhdefs}, leaves us with
\besp
\label{even}
\mathcal{L}_e =  - \frac{1}{2}\partial_\rho h_{\mu\nu}\cu*\partial^\rho h^{\mu\nu}-\frac{s}{2}\partial_\rho h^\parallel_{\mu\nu}\partial^\rho h^{\parallel\mu\nu}
+s\,\partial^\nu h^\parallel_{\mu\nu}\partial_\rho h^{\parallel\mu\rho} +2M^2h^\parallel_{\mu\nu}\cu*h^{\mu\nu}\\
-\frac{1}{2}\partial_\rho h^\perp_{\mu\nu|ab}\cu*\partial^\rho h^{\perp\mu\nu|ab}
+\partial^\nu h^\perp_{\mu\nu|ab}\cu*\partial_\rho h^{\perp\mu\rho|ab}\,.
\eesp
We see from the second line, that the perpendicular components propagate amongst themselves. These kinetic terms are diagonalised by using \eqref{hodgesquared} to define the  (anti)self-dual combinations:
\be 
\label{perpcombo}
h^{\perp\pm}_{\mu\nu|ab} = \frac12(h^\perp_{\mu\nu|ab}\pm*h^\perp_{\mu\nu}{}^{|ab})\,.
\ee
Note that covariant and contravariant $a$-type indices are here identified, so the definition is basis dependent. However it makes it clear that the kinetic terms split into fields that propagate with the right sign, and ghost-like fields that propagate with the wrong sign:
\be 
\label{perps}
-\frac{1}{2}\partial_\rho h^{\perp+}_{\mu\nu|ab}\partial^\rho h^{\perp+\mu\nu}{}_{|ab}
+\partial^\nu h^{\perp+}_{\mu\nu|ab}\partial_\rho h^{\perp+\mu\rho}{}_{|ab} 
+\frac{1}{2}\partial_\rho h^{\perp-}_{\mu\nu|ab}\partial^\rho h^{\perp-\mu\nu}{}_{|ab}
-\partial^\nu h^{\perp-}_{\mu\nu|ab}\partial_\rho h^{\perp-\mu\rho}{}_{|ab}\,,
\ee
although none of these can be regarded as physical since they are massless but all 9 traceless $(\mu\nu)$ polarisations propagate for all $[ab]$. Thus these will all have to gain a regulator mass when spontaneous symmetry breaking is imposed. \TRM{Note that these are a subset of the fields in \eqref{candidates}, that we might have mistaken as graviton degrees of freedom. }

Meanwhile from the rest of  \eqref{even}, $*h_{\mu\nu}$ imposes
\be 
\square h_{\mu\nu} + 4M^2 h_{\mu\nu}^\parallel = 0\,,
\ee
and so we can deduce that since $h_{\mu\nu}$ has been gauge fixed to be transverse  traceless, then $h_{\mu\nu}^\parallel$ is also transverse traceless. Additionally we see that while the rest of $*h_{\mu\nu}$ behaves as a Lagrange multiplier, its  transverse traceless part, $*h^{tt}_{\mu\nu}$, propagates through $h_{\mu\nu}$. Defining
\be 
\label{hpm}
h^\pm_{\mu\nu} = \frac12(h_{\mu\nu}\pm*h^{tt}_{\mu\nu})
\ee
(similar to \eqref{perpcombo} but however this time basis \emph{independent}) we see from \eqref{even} that one of these could be identified with the graviton and one must be a ghost with wrong sign kinetic term:
\be 
\label{threesys}
- \frac{1}{2}\partial_\rho h^+_{\mu\nu}\partial^\rho h^{+\mu\nu}
+ \frac{1}{2}\partial_\rho h^-_{\mu\nu}\partial^\rho h^{-\mu\nu}
-\frac{s}{2}\partial_\rho h^\parallel_{\mu\nu}\partial^\rho h^{\parallel\mu\nu}
 +2M^2h^\parallel_{\mu\nu}(h^{+\mu\nu}-h^{-\mu\nu})\,.
\ee
Which has the right sign depends on the sign chosen for the action: see \eqref{sumL} and the discussion below \eqref{action}. By choosing the sign $s$ of the Pfaffian, we can also ensure that $h^\parallel_{\mu\nu}$ propagates with the right sign. Since it has only the two transverse polarisations, it too could qualify as the graviton. Since we cannot have a theory with two gravitons that self-interact \cite{Boulanger:2000rq}, only one of these two contenders could ultimately play the r\^ole. Which gets chosen will depend on the symmetry breaking mechanism.  

The three fields in \eqref{threesys} are coupled together by what appear to be mass terms. Writing 
$X^T_{\mu\nu}=(h^+_{\mu\nu}, h^-_{\mu\nu}, h^\parallel_{\mu\nu})$, $U^T=(1,-1,0)$, $V^T=(0,0,1)$ and $D=\text{diag}(1, -1, s)$, \eqref{threesys} takes the form
\be 
\frac12\, X^T_{\mu\nu} D\Box X^{\mu\nu} + M^2 X^T_{\mu\nu} A X^{\mu\nu}\,,
\ee
where $A= U V^T+V U^T$. If the kinetic terms were all of the right sign, the mass matrix $2M^2A$ could be diagonalised, but the presence of ghosts prevents this. In fact, setting the normalisation of the action to $-1/\alpha$ (so that in \eqref{action}, $\alpha=1/2$),
 the propagator is 
\be
\label{prop}
\langle X_{\mu\nu}(p) X^{T \rho\sigma}(-p)\rangle = \alpha \, \Pi^{\rho\sigma}_{\mu\nu} \, \Delta\,,
\ee
where $\Pi^{\rho\sigma}_{\mu\nu}$ is the transverse traceless projector on the space of symmetric tensor fields,
and the reduced propagator
\be
\label{Delta}
\Delta = \langle X X^T\rangle = (p^2 D-2M^2A)^{-1} = {D}/{p^2} +2M^2DAD/p^4+4M^4DADAD/p^6\,.
\ee
The expansion in $1/p^2$ terminates because $(DA)^3$, equivalently $(AD)^3$, vanishes. Therefore the mass matrix in \eqref{threesys} does not actually result in masses but rather encodes further propagator-like contributions with improved ultraviolet behaviour ($1/p^4$ and $1/p^6$ respectively). Introducing $W^T=(1,1,0)$, we have $DAD=s(WV^T+VW^T)$ and $DADAD=sWW^T$, and thus read off:
\beal
\label{hprops}
\langle h^+ h^+\rangle &= \frac1{p^2}+4s\frac{M^4}{p^6}\,, & 
\langle h^- h^-\rangle &= -\frac{1}{p^2}+4s\frac{M^4}{p^6}\,, & 
\langle h^\parallel h^\parallel\rangle &= \frac{s}{p^2}\,,\nn\\
\langle h^+ h^-\rangle &= 4s\frac{M^4}{p^6}\,, & 
\langle h^+ h^\parallel\rangle &= 2s\frac{M^2}{p^4}\,, & 
\langle h^- h^\parallel\rangle &= 2s\frac{M^2}{p^4}\,.
\eeal

This behaviour can be understood as follows. Recall that the mass scale $M$ was introduced in \eqref{componentfields} and is arbitrary. Therefore it would make no sense if it resulted in propagators that contain $M$ as a genuine mass. In fact recalling also \eqref{measure}, we can eliminate $M$ by rescaling 
\be 
\label{Maway}
h_{|a_1\cdots a_p}\mapsto M^{-p}\, h_{|a_1\cdots a_p}\,,\qquad \alpha\mapsto M^{-4}\, \alpha\,.
\ee
Since this means that the fields now have differing dimensions $[h]=1$, $[h^\parallel]=3$ and $[*h]=5$, and $M$ is no longer available to fix dimensions, non-vanishing propagators for these fields must have a unique power of $p$ as determined by dimensions, which may thus differ from the standard $1/p^2$. Note that the (anti)self-dual fields \eqref{hpm} no longer have a well defined mass dimension, so to see this effect we need to work in the original basis. From \eqref{Delta} we read off,
\beal
\label{hnomassprops}
\langle h\, h\rangle &= \frac{W^T\!\Delta W}{M^4} =16\frac{s}{p^6}\,, & 
\langle *h \,\cu*h\rangle &=  M^4U^T\!\Delta U = 0 \,, & 
\langle h^\parallel \,h^\parallel\rangle &= V^T\!\Delta V =\frac{s}{p^2}\,,\nn\\
\langle h \,\cu*h^{tt}\rangle &= W^T\!\Delta U = \frac2{p^2}\,, & 
\langle h\, h^\parallel\rangle &= \frac{W^T\!\Delta V}{M^2}  = 4\frac{s}{p^4}\,, & 
\langle *h\, h^\parallel\rangle &= M^2U^T\!\Delta V = 0\,,
\eeal
where a factor of $1/M^4$ is provided by rescaling $\alpha$ in \eqref{prop}.
We see that indeed $M$ has disappeared, and that the propagators are dimensionally correct on recalling that $[\alpha]=4$ now in \eqref{prop}. 

\subsection{Fermionic sector}

Finally we turn to the remaining fermionic fields, \eqref{LoOrig}. Taking the Hodge dual of \eqref{epsrelation} and using \eqref{hodgesquared} and $s^2=1$,\footnote{or note that since \eqref{epsrelation} holds for any $4\cu\times4$ invertible antisymmetric matrix it holds for $\epsilon^{ab}$, and its Pfaffian $1/s$.} we have
\be 
\cu*\epsilon_{ab} := \frac12\epsilon_{abcd}\epsilon^{cd} = s\, \epsilon_{ab}\,.
\ee
Evidently this is also what one gets by lowering indices on \eqref{epsrelation} using \eqref{raiselower}, so the notation is unambiguous.  Noting that $*h^{\mu a}$ now behaves as a Lagrange multiplier imposing transversality (Lorentz gauge), $\partial^\nu h_{\mu\nu|a} = 0$, \eqref{LoOrig} further simplifies (to just three terms). It is however clear that at this stage we should fully split into transverse traceless and transverse traceful degrees of freedom:
\be 
h_{\mu\nu|a}(p) = h^{tt}_{\mu\nu|a}(p)+\frac23\Pi^t_{\mu\nu}(p)\,\varphi_{|a}(p)\,,\qquad
*h^{\mu\nu|a} = *h^{tt\mu\nu|a}+\frac23\Pi^{t\mu\nu}\cu*\varphi^{|a}\,,
\ee
where
\be
\Pi^t_{\mu\nu}(p) = \delta_{\mu\nu} - \frac{p_\mu p_\nu}{p^2}
\ee
is the transverse traceful part (the projector on vector fields into the transverse space), 
the transverse traceless modes satisfy $h^{tt\mu}{}_{\mu|a}=0$ and $*h^{tt}_\mu{}^{\mu|a}=0$, and the coefficient $2/3$ is fixed by taking the trace and comparing with the definition \eqref{phi}. Then in
\be 
\label{Lo}
\mathcal{L}_o =  -\new{\frac12}\partial_\rho h^{tt}_{\mu\nu}{}^{|a}\epsilon_{ab}\,\cu*\partial^\rho h^{tt\mu\nu|b} 
-\frac{s}{2}M^2 \cu*h^{tt}_{\mu\nu}{}^{|a}\epsilon_{ab}\,\cu* h^{tt\mu\nu|b} 
 -\new{\frac23}\partial_\rho \varphi^{|a}\epsilon_{ab}\cu*\partial^\rho \varphi^{|b} 
+ \frac43 sM^2 \cu*\varphi^{|a}\epsilon_{ab}\cu*\varphi^{|b} \,,
\ee
the traceless and traceful modes decouple. These fields are wrong-statistics and thus must all be made to gain a mass via some future symmetry breaking mechanism. Again despite appearances, $M$ does not play this r\^ole. Writing $Y^{Ta}=(\varphi^{|a},*\varphi^{|a})$ and writing the projectors $\sigma_\pm = \half(\one\pm\sigma_3)$, the transverse traceful part of the action is
\be 
\label{Lotrace}
\new{\frac13}\,Y^{Ta}\,\epsilon_{ab}\left( \sigma_1\Box +\new{2}sM^2\sigma_-\right)Y^b\,,
\ee
where the $\sigma_i$ are the Pauli matrices. With normalisation factor above \eqref{prop}, the propagator is then
\be 
\langle Y^a(p) Y^{Tb}(-p)\rangle = -\frac{3\alpha}{\new{2}}\epsilon^{ab} (p^2\sigma_1-\new{2}sM^2\sigma_-)^{-1} = -\frac{3\alpha}{\new2} \epsilon^{ab} \left( \frac{\sigma_1}{p^2}+\new{2}sM^2\frac{\sigma_+}{p^4}\right)\,,
\ee
the expansion in $1/p^2$ terminating because $\sigma_1\sigma_-\sigma_1\sigma_-\sigma_1 = \sigma_+\sigma_-\sigma_1 = 0$. We thus see that 
\be 
\label{scalarfermion}
\langle\varphi^{|a}\, \varphi^{|b}\rangle = -\new{3}\alpha sM^2\frac{\epsilon^{ab}}{p^4}\,,\qquad
\langle *\varphi^{|a}\, \cu*\varphi^{|b}\rangle = 0\,,\qquad
\langle\varphi^{|a} \,\cu*\varphi^{|b}\rangle =-\frac{3\alpha}{\new2}\frac{\epsilon^{ab}}{p^2}\,.
\ee
We see again the same effect: the mass term in \eqref{Lotrace} does not actually behave as a mass but rather provides propagators with improved ultraviolet behaviour. Again this can be understood by dimensions and by the fact that $M$ is arbitrary. Indeed we see that the transformation \eqref{Maway} removes all reference to $M$. 

Since from \eqref{Lo}, the pattern is the same for the transverse traceless modes, we have immediately that
\be 
\label{tensorfermion}
\langle h^{tt}_{\mu\nu}{}^{|a}\, h^{tt\rho\sigma|b}\rangle = \alpha sM^2\Pi^{\rho\sigma}_{\mu\nu}\frac{\epsilon^{ab}}{p^4}\,,\quad
\langle *h^{tt}_{\mu\nu}{}^{|a}\, \cu*h^{tt\rho\sigma|b}\rangle = 0\,,\quad
\langle h^{tt}_{\mu\nu}{}^{|a}\, \cu*h^{tt\rho\sigma|b}\rangle =-\new{2}\alpha\Pi^{\rho\sigma}_{\mu\nu}\frac{\epsilon^{ab}}{p^2}\,.
\ee

\section{Summary and discussion}
\label{sec:discussion}

As reviewed in sec. \ref{sec:intro},  the Parisi-Sourlas regularisation works in gauge theory by adding to the original gauge field $A^1_\mu$,  a complex pair of fermionic gauge fields $B_\mu, \bar{B}_\mu$ and a ghost copy, $A^2_\mu$. For the pure $SU(N|N)$ gauge theory at the free level, only the two transverse polarisations propagate for all these fields and they are decoupled from each other. 

In the analogous situation in gravity, the solution already at the free level is much richer and more subtle. We have to expand around a non-vanishing background, \eqref{flatg}, which in a sense already leads to some spontaneous symmetry breaking. However the resulting mass-like terms do not actually provide masses but are responsible for providing further propagators with improved ultraviolet behaviour (\viz $1/p^4$ and $1/p^6$). 

The propagating modes even at the free level are not just transverse traceless ones, as expected for the graviton. The transverse traceless bosonic modes form a multiplet,  $h_{\mu\nu}(x)$, $*h^{tt}_{\mu\nu}(x)$ and $h^\parallel_{\mu\nu}(x)$, that  propagate into each other through $1/p^2$ and higher powers, according to \eqref{hnomassprops}. Defining (anti)self-dual combinations \eqref{hpm} out of the first pair, one of $h^\pm_{\mu\nu}$ is a ghost, while the other has the right sign propagator, \cf \eqref{threesys} and \eqref{hprops}. We also saw that $h^\parallel_{\mu\nu}$ can have either sign propagator depending on the sign of $s$ (the Pfaffian of the fermionic part $\epsilon_{ab}$ of the flat metric). $h^\parallel_{\mu\nu}$  is the part of $h_{\mu\nu|ab}(x)$ that is parallel to $\epsilon_{ab}$. As we saw in \eqref{perps}, the perpendicular part, $h^{\perp\pm}_{\mu\nu|ab}$, is traceless but not transverse. They do not mix; one propagates as a real field while the other propagates as a ghost-field \cf \eqref{perps}.


This summarises all the bosonic propagating modes. The fermionic modes $h_{\mu\nu|a}$ and $*h^{\mu\nu|a}$ are wrong-statistics fields, thus intended to be Pauli-Villars. They are all transverse but split into transverse traceless and transverse traceful. Each of these form a doublet propagating into each other with $1/p^2$ and $1/p^4$ propagators, according to \eqref{tensorfermion} and \eqref{scalarfermion} respectively.
 
Turning to the non-propagating modes, all of the $h_{ab}(x,\theta)$ superfield is eliminated algebraically via the linearised superdiffeomorphisms \eqref{habgauge}, except for $*h^{ab}(x)$. This latter is gauge invariant but becomes a Lagrange multiplier enforcing the tracelessness of $h_{\mu\nu|ab}(x)$ (on its first two indices). Similarly all of $h_{\mu a}(x,\theta)$ can be gauged away, apart from $*h^{\mu a}(x)$ which is gauge invariant but behaves as a Lagrange multiplier imposing transversality of the propagating fermionic modes. A remaining $\xi'_a(x)$ gauge invariance allows to impose the radiation gauge $\varphi_{|a}=0$, \cf \eqref{xipha}, while the original bosonic gauge invariance carried by $\xi_\mu(x)$ allows to choose $h_{\mu\nu}(x)$ to be transverse traceless. 

Finally, the vector field $h_{\mu(a|b)}(x)$ is special in that it and its gauge invariance, \eqref{hmuabtrans}, are untouched and absent from the free action. At the interacting level, it could act as a Lagrange multiplier leading to constraints on the form of the allowed spontaneous symmetry breaking. 

Finding such a symmetry breaking is the next most important step in this construction. One can expect to need to induce all modes to gain a mass, apart from the graviton,  analogous to that achieved for $U(1|1)$ gauge theory in ref. \cite{Falls:2017nnu}, since the kind of decoupling otherwise seen in gauge theory, \cf sec. \ref{sec:intro} \cite{Arnone:2001iy}, is unlikely to be effective here. In view of the similarity of a cosmological constant term to a mass-term for the graviton when expanded around flat space, this seems a promising starting point.  Of course in normal bosonic (Einstein) gravity, a cosmological constant does not provide a mass-term since diffeomorphism invariance is still unbroken, and the linearised part $\propto \kappa \varphi$, \cf \eqref{cc} and \eqref{phi}, is anyway more important, signalling that flat space is no longer a classical solution. Here however we see from \eqref{cc}  that the cosmological constant will induce curvature only in $*h_{\mu\nu}(x)$ and $*h_{ab}(x)$ to first order  (since by \eqref{measure} and \eqref{hodge} only these components have non-vanishing integrals to first order in $\kappa$). In the cosmological constant term, $h^\parallel_{\mu\nu}$ appears first only at second order where, thanks to \eqref{s}, it takes the form of a mass-term.

\TRM{The properties of this Parisi-Sourlas supergravity construction already clearly differ from standard realisations of supergravity. We highlight where these differences enter and compare to other extensions of supergravity. In standard ($N=1$, $D=4$) supergravity  there are also four fermionic coordinates but they are cast as a complex conjugate pair of two-component coordinates $\theta^\alpha$ and $\bar\theta^{\dot\alpha}$. Most importantly we  set the torsion field to vanish, in order for the regularising structure to maintain the close similarity to the graviton interactions in the Einstein-Hilbert action. In the standard realisation of supergravity the torsion field is non-vanishing even in flat space, being related to the Pauli matrices $\sigma^\mu_{\alpha\dot\alpha} \sim (i,\bm{\sigma})$, and the tangent space symmetry of $\theta^\alpha$ and $\bar\theta^{\dot\alpha}$ is then tied to the bosonic vectorial Lorentz representation, see \eg \cite{Gates:1983nr}. The Parisi-Sourlas supergravity developed here could  therefore be viewed as a kind of deformation of standard supergravity. Since expansion over the $\theta^a$ leads to component  fields carrying antisymmetric vectorial indices (the fermionic $a,b,\cdots$) reminiscent of forms, and thus also leading to fields with mixed representations,
it has some superficial resemblance to Generalized Geometry \cite{Hitchin:2004ut,Gualtieri:2003dx}. However the indices $a,b,\cdots$ are not associated to the cotangent bundle but belong to a new space. 
This latter property gives the theory also an apparent resemblance to Double Field Theory \cite{Siegel:1993th,Hull:2009mi}, although there is no doubling of the bosonic coordinates here or relation to $T$-duality.} 

\section*{Acknowledgments}
MPK and TRM acknowledge support via an STFC PhD studentship and Consolidated Grant ST/P000711/1 respectively.

\bibliographystyle{hunsrt}
\bibliography{references} 

\end{document}